# Metal-bonded Atomic Layers of Transition Metal Carbides (MXenes)

*Zongju Cheng[#], Zhiguo Du[#], Hao Chen, Qi Zhao, Yu Shi, Haiyang Wang, Yuxuan Ye and Shubin Yang\**

Dr. Z. J. Cheng, Dr. Z. G. Du, Dr. H. Chen, Dr. Q. Zhao, Dr. Y. Shi, Dr. H. Y. Wang, Dr. Y. X. Ye and Prof. S. B. Yang

School of Materials Science and Engineering, Beihang University, 100191, Beijing, China

[#]These authors contribute equally to this work.

*Email: yangshubin@buaa.edu.cn

**Although two-dimensional transition metal carbides and nitrides (MXenes) have fantastic physical and chemical properties as well as wide applications, it remains challenging to produce stable MXenes due to their rapid structural degradation. Here, unique metal-bonded atomic layers of transition metal carbides with high stabilities are produced via a simple topological reaction between chlorine-terminated MXenes and selected metals, where the metals enable to not only remove Cl terminations, but also efficiently bond with adjacent atomic MXene slabs, driven by the symmetry of MAX phases. The films constructed from Al-bonded $Ti_3C_2Cl_x$ atomic layers show high oxidation resistance up to 400 °C and low sheet resistance of 9.3 Ω/□. Coupled to the multi-layer structure, the Al-bonded $Ti_3C_2Cl_x$ film displays a significantly improved EMI shielding capability with a total shielding effectiveness value of 39 dB at a low thickness of 3.1 μm, outperforming pure $Ti_3C_2Cl_x$ film.**

The rapid development of integrated electronic devices and wide applications of the telecommunication networks have caused serious electromagnetic interference (EMI), which has detrimental influence on functionality of electronics and even human health.[1–5] Therefore, it is of urgent demand to explore suitable shielding materials to efficiently absorb or reflect the undesirable electromagnetic radiation. According to Schelkunoffs theory, the total EMI shielding effectiveness ($SE_T$) is comprised of three parts: (1) reflecting attenuation ($SE_R$), originated from impedance mismatch between the propagation mediums; (2) absorbing attenuation ($SE_A$), depending on electrical conductivity and thickness of the EMI shields; (3) multiple reflection attenuation ($SE_M$), being negligible as the total shielding effectiveness is >15 dB.[6] Thus, high-electrical-conductive metals, such as Al, Cu, Ag, have been widely utilized for EMI shielding.[7–10] However, these metals commonly possess high densities (2.7-10.5 g·cm$^{-3}$) and poor processability, rendering them difficult to be used in tiny mobile electronic devices. Compared with the metals, carbonaceous materials (carbon nanotube, graphene) and carbides/nitrides (MXenes), possessing lower densities (0.3-2.6 g·cm$^{-3}$), have been investigated as promising candidates for EMI shielding.[11–17] In particular, $Ti_3C_2T_x$ MXene films commonly exhibit high EMI shielding capabilities (29-92 dB), determinate to the film thickness.[18,19]

MXenes are atomic layers of carbides, nitrides and carbonitrides with a general formula of $M_{n+1}X_nT_x$ (n=1-4), where M is early transition metals (e.g. Ti, V, Nb), X is carbon and/or nitrogen and T refers to the surface terminations (e.g. -F, -O, -OH, -Cl).[20,21] To date, MXenes have been synthesized by extracting A species from MAX phases in fluorine-containing acid solution,[22–27] molten salts[28,29] and alkali solution.[30,31] Among these methods, molten salts etching method is unique to engineering surface terminations of MXenes by using different halide salts (e.g. $CuCl_2$, $CuBr_2$, CuI).[28,32] Moreover, Cl and Br terminations can be further substituted with -S, -Se, -Te, -NH by chemical reaction between halogen-functionalized MXenes and corresponding active molten salts.[33] Interestingly, termination-substituted $Nb_2CT_x$ (T= -S, -Se, -NH) exhibited unexpected superconductivity. In addition, Cl-terminated MXenes show a great stability up to 750 °C without structural degradation under Ar atmosphere, which is in contrast to the MXenes produced via liquid etching that show poor stability.[34–37] Unfortunately, these halogen-functionalized MXenes have relatively high electrical resistances (>$10^4$ Ω/□), limiting their EMI shielding application.

Here, we developed an efficient strategy to synthesize unique metal-bonded MXenes by a topological reaction between Cl-terminated MXene and selected metal species at high temperatures (700-900 °C). During the reaction process, metal-bonded MXenes would be gradually formed between adjacent layers driven by the suitable crystal symmetry of MAX phases, associated with the removal of Cl terminations by the gaseous by-products. The exfoliated MXene layers can be further transformed into metal-bonded MXene nanosheets (such as Al-$Ti_3C_2Cl_x$), which can be employed as building blocks to construct a large area and continuous films, displaying a good oxidation resistance up to 400 °C and low sheet resistance of 9.3 Ω/□, which is three magnitude lower than that of $Ti_3C_2Cl_x$ films ($4.0\times10^4$ Ω/□). Benefited from the multi-interface structure and low sheet resistance, the resultant film exhibits a good EMI shielding performance with a total SE value of 39 dB at a low thickness of 3.1 µm.

Metal-bonded MXenes were synthesized through annealing treatment of chlorine-terminated MXenes and metal powders at a high temperature of 700-900 °C (see Experiment Section in Supporting Information). Taking Al-bonded $Ti_3C_2Cl_x$ MXene (Al-$Ti_3C_2Cl_x$) as an example, during the reaction process, the metal Al would be reacted with the Cl terminations of $Ti_3C_2Cl_x$ MXene (Figure S1 and S2, Supporting Information) to form gaseous aluminum chloride ($AlCl_3$, $T_{sublimating}$=180 °C) owing to the higher bonding energy of Al-Cl (507±1.0 kJ/mol) than that of Ti-Cl (405.4±10.5 kJ/mol).[33] The by-product gaseous aluminum chloride would be facilely removed from the reaction system (Table S2, Supporting Information). Simultaneously, extra metal Al would react with exposed Ti atoms and zip the adjacent layers of $Ti_3C_2Cl_x$ MXene, generating Al-bonded Ti$_3$C$_2$Cl$_x$ MXene. After the reaction, the black MXene powder would be transformed to grey powder (Figure S3, Supporting Information).

To characterize the morphology and structure of Al-Ti$_3$C$_2$Cl$_x$ MXene, transmission electron microscopy (TEM) and High-resolution TEM (HRTEM) were conducted. As shown in **Figure 1**, in the case of Al-Ti$_3$C$_2$Cl$_x$ MXene, an accordion-like morphology is observed, containing numerous slender straps with few to tens of nanometers width (Figure S4, Supporting Information), perfectly inheriting to the parent Ti$_3$C$_2$Cl$_x$ MXene (Figure 1b and Figure S5, Supporting Information). The cross-section image reveals that the adjacent straps of Al-Ti$_3$C$_2$Cl$_x$ MXene became closer and more straight compared with the initial Ti$_3$C$_2$Cl$_x$ MXene. Atomic-resolution high angle annular dark field scanning transmission electron microscope (HAADF-STEM) images reveals that, in the marked section (Figure 1f and Figure S6, Supporting Information), single atomic align is well-defined immobilized between two adjacent slabs in Al-Ti$_3$C$_2$Cl$_x$ MXene, whereas the previous two layers of Cl terminations in Ti$_3$C$_2$Cl$_x$ MXene layers are invisible (Figure 1c and Figure S7, Supporting Information). Enlarged atomic image of Al-Ti$_3$C$_2$Cl$_x$ MXene clearly exhibits that two adjacent slabs composed of three bright atomic layers are connected by one dark atomic layer. In STEM images, the brightness of atoms is proportional to the square of their atomic number. Thus, the dark and bright layers of Al-Ti$_3$C$_2$Cl$_x$ MXene are Al and Ti layers, respectively, whose model with golden and red balls are inserted in Figure 1g. The interlayer spacing of Al-Ti$_3$C$_2$Cl$_x$ MXene is measured to be 9.45 Å, which is much lower than that of Ti$_3$C$_2$Cl$_x$ layers (11.47 Å) in Figure 1d, close to that of Ti$_3$AlC$_2$ MAX phase (9.21 Å). This shrinkage of interlayer spacing should be driven by the suitable crystal symmetry of MAX phases. Unexpectedly, in the exterior of one Al-bonded Ti$_3$C$_2$Cl$_x$ MXene slab, small and bright atoms are aligned on the outmost Ti atoms, which should be Cl terminations. This suggests that Al atoms enable to bond adjacent Ti$_3$C$_2$Cl$_x$ MXene slabs, rather than substitute the outmost Cl terminations. Thus, partly Al-bonded MXene and the unique heterostructures are achieved.

To further gain insight into the reaction process, the ex-situ X-ray diffraction (XRD) measurement was carried out. As shown in **Figure 2**a, the pristine mixture ($Ti_3C_2Cl_x$ and Al) exhibits a series of sharp peaks at 7.9°, 15.9°, 24.0°, 57.8° and 38.4°, 44.6°, 65.0°, indexed to (002), (004), (006) and (110) facets of $Ti_3C_2Cl_x$ MXene and (111), (200) and (220) facets of Al metal (marked with black inverted triangles), respectively. With increasing the annealing times from 0 to 5 min, the intensities of three Al peaks are gradually decreased, and become indiscernible after 3 min. Meanwhile, the (002) peak of $Ti_3C_2Cl_x$ MXene was continuously broadened towards higher degree, indicating the interlayer spacing shrinkage. This shrinkage was terminated after annealing for 5 min, associated with a new peak presented at 9.2°, indexed to the (002) facet of Al-$Ti_3C_2Cl_x$ MXene. This value is higher than that of initial $Ti_3C_2Cl_x$ MXene (7.9 °), close to that of $Ti_3AlC_2$ MAX phase (9.4°). The calculated (002) interplanar spacing is 9.58 Å, in good agreement with the observation from above HAADF-STEM image (Figure 1g). Moreover, another dominated peak is presented at 38.8°, much stronger and higher than that of $Ti_3C_2Cl_x$ MXene (38.4°), close to (104) facet of $Ti_3AlC_2$ MAX phase (39.0°). This should be originated from the presence of a crystal symmetry of crystalline $Ti_3AlC_2$ MAX phase in $Ti_3C_2Cl_x$ MXene. Except for the peaks of Al-$Ti_3C_2Cl_x$ and $Ti_3C_2Cl_x$ MXene, no obvious peak of impurities (such as Ti-Al alloy) is detected, suggesting that Al species was only participated in bonding reaction. Remarkably, the formation of Al-$Ti_3C_2Cl_x$ MXene should be initiated within 3 min, verified by the occurrence of the white smoke in the tube furnace, which is identified as $AlCl_3$ (Figure S8, Supporting Information). Such rapid reaction discloses the thermodynamics feasibility and fast kinetics of this metal bonding reaction. In this principle, other metals should be also bonded with Cl-MXenes. Thus, different metal elements (Si, Ga, Ge, In, Sn) were annealed with $Ti_3C_2Cl_x$ MXene at the temperature of 800-900 °C (details in Table S1, Supporting Information). The resultants possess a relatively weak peak adjacent to the (002) peak of $Ti_3C_2Cl_x$ MXene in Figure 2b, similar to Al-bonded $Ti_3C_2Cl_x$ MXene, clearly demonstrating the successful synthesis of metal-bonded $Ti_3C_2Cl_x$ MXene. As-produced metal-bonded $Ti_3C_2Cl_x$ MXenes have accordion-like morphology with homogeneous distribution of Ti, Cl and corresponding metal elements (Figure S9-13, Supporting Information). Remarkably, as reaction with Si and Ge powders, they were maintaining solid under reacting temperature (Table S2, Supporting Information), indicating that solid diffusion reaction is also feasible for this topological reaction. Therefore, more metals with high melting points could be possibly induced to produce metal-bonded MXenes. As prolonging the annealing time up to 20 min, MAX phases can be even explored (Figure S14 and S15, Supporting Information), disclosing that this metal bonding reaction is the reverse reaction of the selective etching of MAX phases. Thus, it provides a new pathway to synthesize MAX phases, especially for MAX phases that are hardly synthesized from traditional metallurgy sintering, like $Ti_3InC_2$ phase (Figure S16, Supporting Information).[38]

Due to the accordion-like structure and weak interaction between layers (**Figure 3**a and Figure S17, Supporting Information), it is feasible to fabricate Al-$Ti_3C_2Cl_x$ MXene multi-layers with few hundred nanometers in lateral size, in which Ti, Al, Cl and C elements are uniformly dispersed in the atomic layers (Figure S18, Supporting Information). Compared with accordion-like Al-$Ti_3C_2Cl_x$ MXene, multi-layers possess a broadened and weak peak at 9.2° owing to their small size and low thickness (Figure 3b). The as-prepared Al-$Ti_3C_2Cl_x$ nanosheets possess a good crystallinity with an average lattice distance of 0.26 nm, indexed to (100) plane (Figure 3e). Ti atoms in the Al-$Ti_3C_2Cl_x$ nanosheets are regularly arranged of hexagonal symmetry, which is also confirmed by corresponding FFT pattern with six hexagonal diffraction spots (insert, Figure 3e), same to the reported MXenes.[39] The Al-$Ti_3C_2Cl_x$ nanosheets can be further used as building blocks to construct compact films (Figure 3c and 3d, Figure S19, Supporting Information). The cross-section image shows that Al-$Ti_3C_2Cl_x$ film has a thickness of 13 μm (Figure 3c) and homogenous distribution of Ti and Al species in the layers (Figure 3f).

To explore the thermal stability and electrical conductivity of as-prepared Al-$Ti_3C_2Cl_x$ films, thermogravimetry (TG) under air atmosphere and sheet resistance measurements were conducted. As shown in **Figure 4**a, the oxidation process of the Al-$Ti_3C_2Cl_x$ film can be divided into two stages: (1) in the temperature range from 30 °C to 400 °C, the Al-$Ti_3C_2Cl_x$ film remains stable without obvious weight gain; (2) in the temperature range from 400 °C to 800 °C, the film is gradually oxidized associated with a weight gain of 8%. In contrast, the $Ti_3C_2Cl_x$ film is oxidized at a low temperature of 280 °C and completely oxidized at 510 °C. Clearly, the Al-$Ti_3C_2Cl_x$ films possess better oxidation resistance with a higher oxidation temperature and lower oxidation speed than that of $Ti_3C_2Cl_x$ film, which can be attributed to the in-situ formed Ti and Al oxides that prevent oxygen from invading the inner film, as we expected.[40] More importantly, Al-$Ti_3C_2Cl_x$ film has a low sheet resistance of 9.3 Ω/□, substantial lower than that of $Ti_3C_2Cl_x$ ($4.0\times10^4$ Ω/□) (Figure 4b). This electrical conductivity enhancement can be ascribed to the presence of the Ti-Al-Ti configuration that enables to connect individual slabs together. Based on the relatively high electrical conductivity and good oxidation resistance, we preliminarily evaluated the EMI shielding performance of Al-$Ti_3C_2Cl_x$ films in X-band range (8.2−12.4 GHz). As shown in Figure 4d, Al-$Ti_3C_2Cl_x$ film has a total SE value of 31.5 dB, whereas pristine $Ti_3C_2Cl_x$ film displays almost zero EMI shielding effectiveness that is close to the value of bare mica substrate (Figure S20, Supporting Information). To enhance the shielding performance, Al-$Ti_3C_2Cl_x$ films with increased thickness were fabricated, as shown in Figure 4e. The EMI $SE_T$ values are up to 28.8 dB, 32.5 dB and 39 dB with the film thicknesses of 1.8 μm, 2.3 μm and 3.1 μm, respectively. These values are comparable to those of the HF-produced $Ti_3C_2T_x$ films (Table S3, Supporting Information). This enhancement with increasing thickness is related to the increased multiple interfaces and enhanced electrical conductivity of the Al-$Ti_3C_2Cl_x$ films. As illustrated in Figure 4c, owing to layer-by-layer stacking structure, the films provide many interfaces for incident electromagnetic waves (EM) to undergo the internal reflection, which would enhance the interacting probability between EM and electric dipoles and thus increase the absorbing loss.[41] This is evidenced by variations of the absolute values of $SE_T$, $SE_R$ and $SE_A$, summarized in Figure 4f. The $SE_A$ values are rapidly rising with increased thickness ($\Delta SE_A$= 8.4 dB), whereas $SE_R$ values are stable and barely dependent of film thickness ($\Delta SE_R$= 1.7 dB). Thus, it can be concluded that EMI shielding enhancement mainly originates from the improved absorption capacity of the films, which is derived from the increase of the multi-interlayer reflections and absorptions of EM waves.

In conclusion, unique metal-bonded MXenes with high stabilities were synthesized through a simple topological reaction between Cl-terminated MXenes and selected metals. During the reaction process, metal A enables to efficiently bond with adjacent MXene layers to form metal-bonded MXene, associated with removal of Cl terminations by generating gaseous metal chlorides. Furthermore, the Al-bonded MXene multi-layers can be further fabricated and assembled into continuous films, which possess the good oxidation resistance and improved electrical conductivity. Benefiting from these merits, Al-$Ti_3C_2Cl_x$ MXene films possess an enhanced EMI shielding performance with the $SE_T$ value of 39 dB in the thickness of 3.1 µm. Combining two-dimensional structures and good properties of metal-bonded MXenes, it would hold great potential in energy storage, catalysis, EMI shielding and radiation protection.

**Supporting Information**

Supporting Information is available from the author.

**Acknowledgements**

This work was financially supported by National Natural Science Foundation of China (Grant Nos. 52125207, 52072014 and 52202204), Beijing Natural Science Foundation (Grant No. JQ20011) and National Postdoctoral Program for Innovative Talents (Grant No. BX20200027).

Received: ((will be filled in by the editorial staff))
Revised: ((will be filled in by the editorial staff))
Published online: ((will be filled in by the editorial staff))

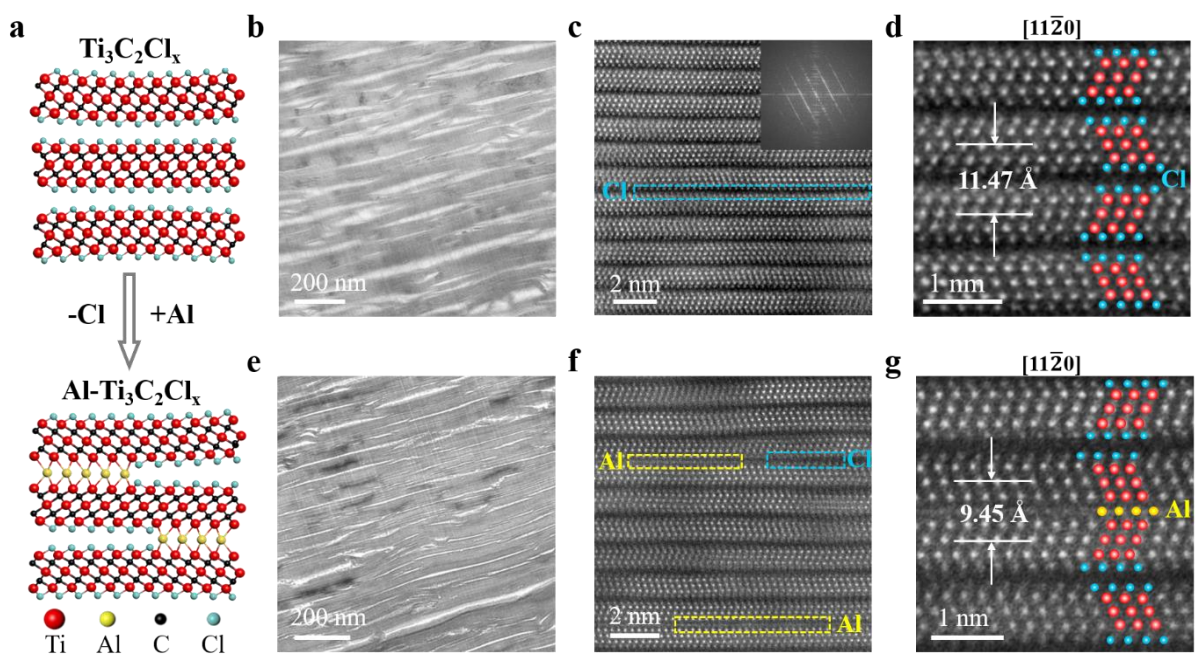

**Figure 1.** Morphology and structure characterization of Al-bonded $Ti_3C_2Cl_x$ MXene. a) Schematics of the atomic structure for $Ti_3C_2Cl_x$ MXene before and after the topological reaction with metal Al. b) TEM image of initial $Ti_3C_2Cl_x$ MXene, exhibiting an accordion-like morphology. c) HAADF and d) enlarged HAADF images, showing that $Ti_3C_2Cl_x$ MXene layers are in ordered arrangement. The inset in c) is the corresponding FFT pattern. e) TEM image of Al-$Ti_3C_2Cl_x$ MXene, exhibiting an accordion-like morphology, well inheriting from $Ti_3C_2Cl_x$ MXene. f) HAADF and g) enlarged HAADF images, showing that single Al layer is immobilized between adjacent $Ti_3C_2Cl_x$ MXene slabs instead of Cl terminations. Ti, Cl and Al atoms are represented by red, cyan and golden balls, respectively.

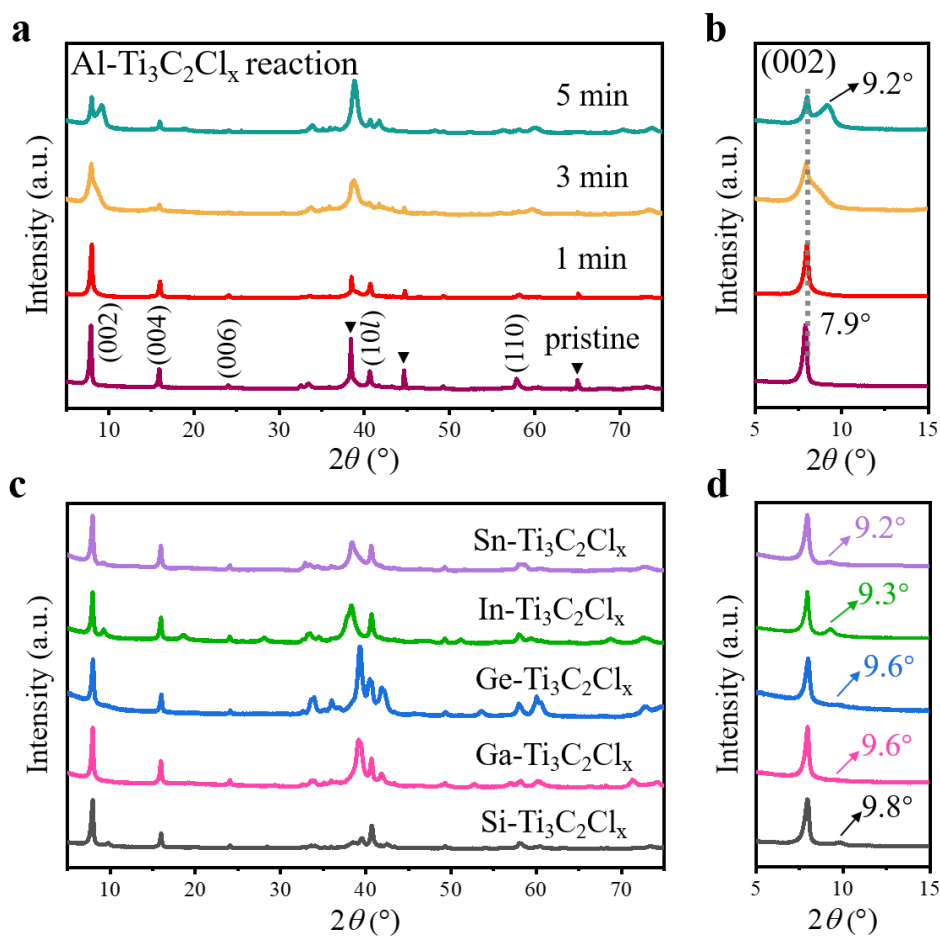

**Figure 2.** Structure characterization of Al-Ti$_3$C$_2$Cl$_x$ during the reaction and different metal-bonded Ti$_3$C$_2$Cl$_x$. a, b) ex-situ XRD patterns (a) and zoom of XRD patterns from 5° to 20° (b) of resultants at the reaction time from 0 to 5 min, exhibiting a new peak at 9.2° adjacent to (002) peak of Ti$_3$C$_2$Cl$_x$ MXene. The peaks marked with black triangle belong to Al powder. c, d) XRD patterns (c) and zoom of XRD patterns from 5° to 20° (d) of different metal-bonded Ti$_3$C$_2$Cl$_x$ MXenes by using various element powders as reactant. All as-produced metal-bonded Ti$_3$C$_2$Cl$_x$ MXenes possess a new peak adjacent to (002) peak of Ti$_3$C$_2$Cl$_x$ MXene, similar to Al-Ti$_3$C$_2$Cl$_x$ MXene.

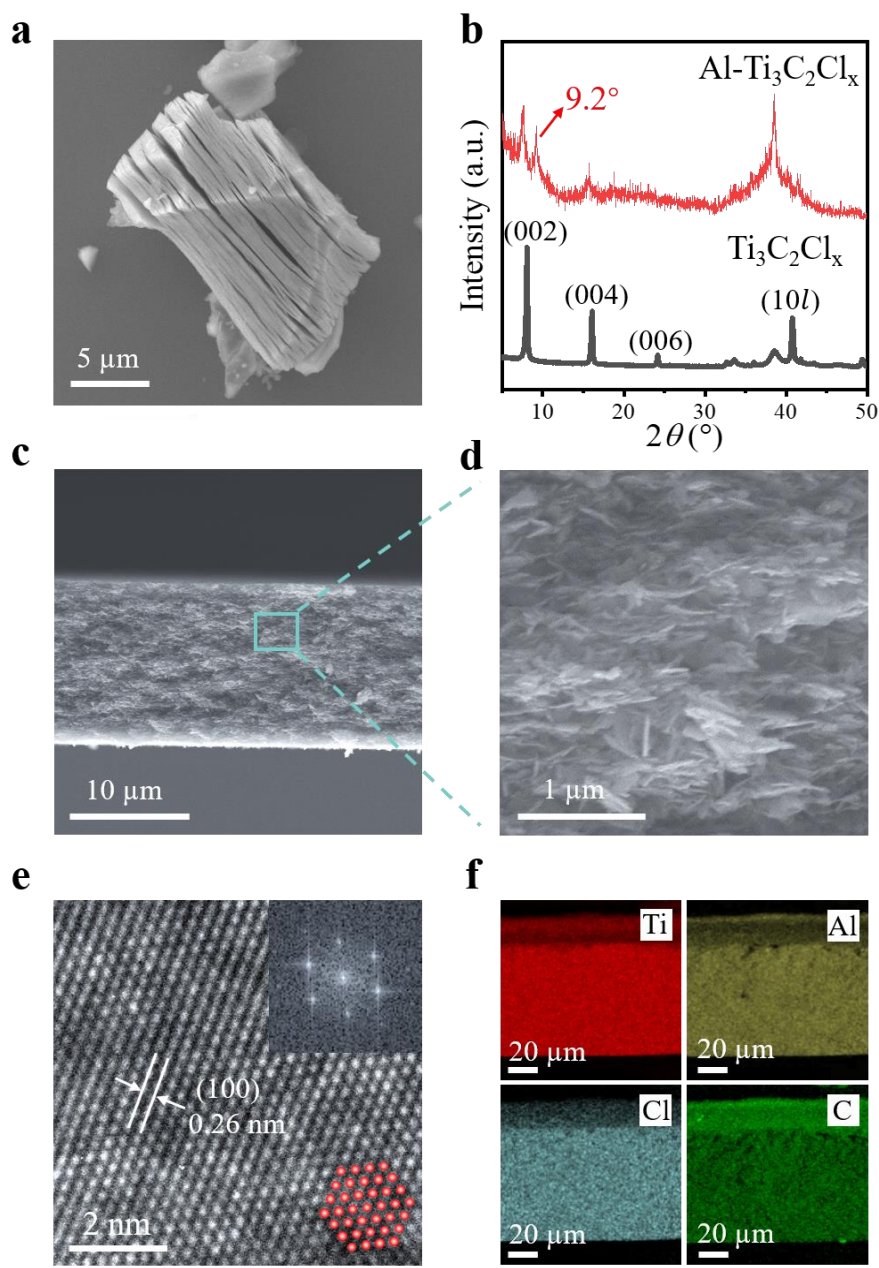

**Figure 3.** Morphology characterization of delaminated Al-bonded $Ti_3C_2Cl_x$ layers and assembled film. a) SEM image of Al-$Ti_3C_2Cl_x$, showing accordion-like morphology. b) XRD patterns of Al-$Ti_3C_2Cl_x$ film, showing an obvious peak at 9.2°. c) Cross-section SEM and d) enlarged images, revealing that the film is densely assembled of small flakes. e) HRTEM and the corresponding FFT (inset) images of Al-$Ti_3C_2Cl_x$ MXene layers, showing a single crystalline feature. f) EDS mapping plots of Al-$Ti_3C_2Cl_x$ film show a homogeneous distribution of Ti, Al, Cl and C species.

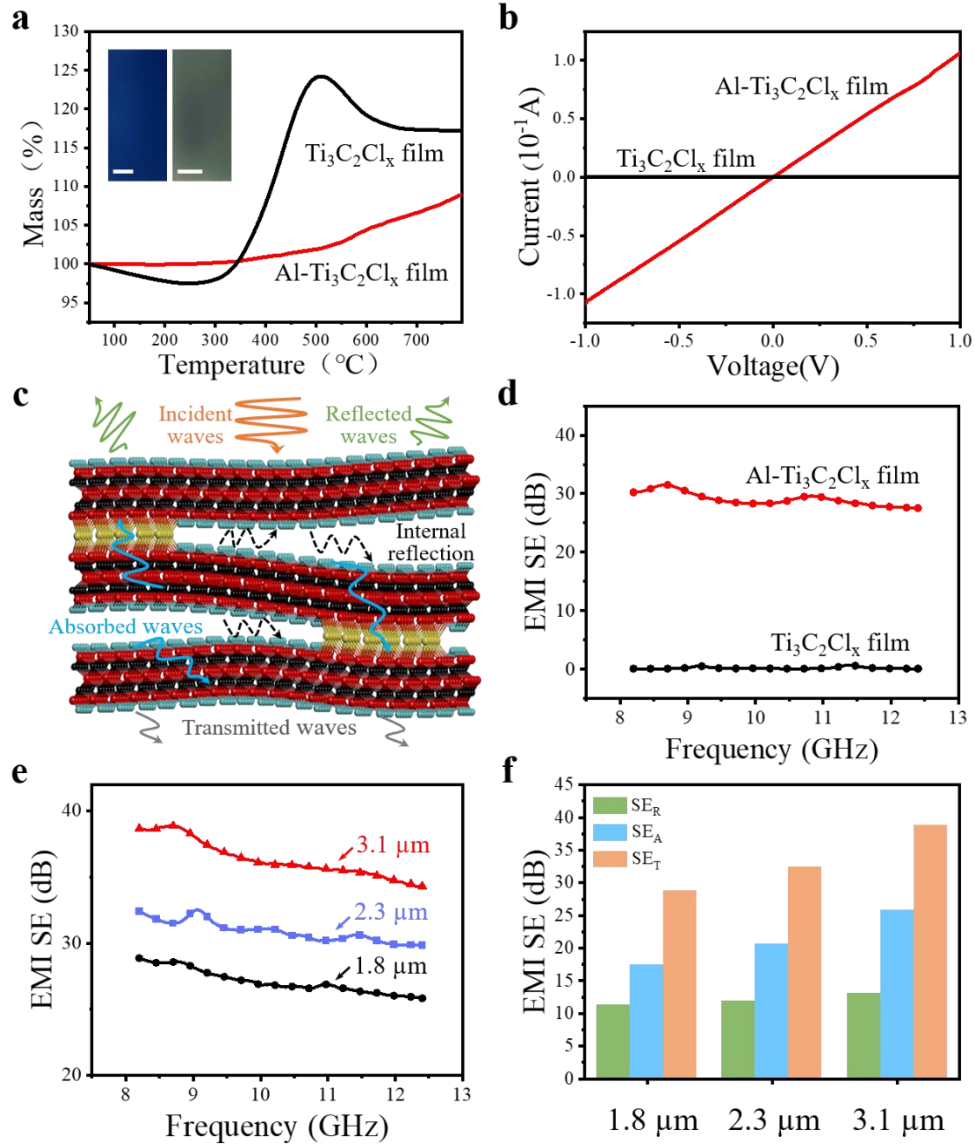

**Figure 4.** Properties and EMI shielding performance of assembled Al-Ti$_3$C$_2$Cl$_x$ films. a) TG curves of Ti$_3$C$_2$Cl$_x$ and Al-Ti$_3$C$_2$Cl$_x$ films. The inset pictures in (a) show that the Al-Ti$_3$C$_2$Cl$_x$ film is grey whereas Ti$_3$C$_2$Cl$_x$ film is blue. scale bar, 0.5 cm. b) I-V curves of films exhibit that Al-Ti$_3$C$_2$Cl$_x$ films possess much lower sheet resistance than Ti$_3$C$_2$Cl$_x$ films. c) Schematic plot of multiple reflections and absorptions in the films. The orange, green, black, blue and grey arrows represent incident, reflected, multi-reflected, absorbed and transmitted EM waves, respectively. d) EMI shielding plots reveal that Al-Ti$_3$C$_2$Cl$_x$ film has a larger SE$_T$ value than Ti$_3$C$_2$Cl$_x$ film. e) EMI SE$_T$ plots of Al-Ti$_3$C$_2$Cl$_x$ films showing an enhanced performance with increased thickness. f) EMI $SE_T$, $SE_R$, and $SE_A$ values of Al-Ti$_3$C$_2$Cl$_x$ films with increased thickness.